\documentclass[prd, twocolumn, showpacs, amsmath, amssymb]{revtex4}%

\usepackage{amsmath}
\usepackage{amsfonts}
\usepackage{amssymb}
\usepackage{graphicx}
\usepackage{epsf,color,colordvi,pifont}
\usepackage{amsfonts}%

\newcommand{\bb}{{\bf b}}

\begin{document}
\title{Holographic duals to Poisson sigma models and noncommutative quantum mechanics}
\author{D. V. Vassilevich}
\affiliation{CMCC, Universidade Federal do ABC, Santo Andr\'e, 0910-580, S.P., Brazil}
\affiliation{Physics Department, St.~Petersburg State University, 198504, St.~Petersburg, Russia}
\begin{abstract}
Poisson sigma models are a very rich class of two-dimensional theories that includes,
in particular, all 2D dilaton gravities. By using the Hamiltonian reduction method,
we show that a Poisson sigma model (with a sufficiently well-behaving Poisson tensor)
on a finite cylinder is equivalent to a noncommutative quantum mechanics for the boundary
data.
\end{abstract}
\pacs{02.40.Gh, 04.62.+v}
\maketitle

\section{Introduction}
The holographic principle \cite{hologram} implies that some quantum theories in $n+1$ dimensions
may be fully or partially equivalent to other theories living in $n$ dimensions. One of the manifestations
of this principle is the celebrated AdS$_{n+1}$/CFT$_n$ correspondence \cite{Maldacena:1997re}.
One may think that the case $n=1$ is the simplest one. The asymptotic symmetry algebras and their
central charges in some AdS$_2$ gravities were analyzed in, e.g., \cite{AdS2cc}.
A conformal quantum mechanics candidate for the holographic theory was discussed in \cite{AdS2QM}. However,
many conceptual problems remain unsolved. It is not even clear yet whether the dual theory should be a
conformal quantum mechanics or a chiral part of a CFT (or, perhaps, both, depending on the model and
boundary conditions). The two-dimensional holography definitely deserves more attention, specifically 
since one may expect some exact results there.

Here we shall consider the Poisson sigma models (PSMs) \cite{Schaller:1994es} in 2D that are described by the action
\begin{equation}
S=\int_{\mathcal{M}} d^2\sigma \epsilon^{\mu\nu} \left[ X^I\partial_\mu A_{\nu I} +\frac 12 P^{IJ}A_{\mu I}A_{\nu J}
\right]\,,\label{PSM}
\end{equation}
where the target space is an $n$-dimensional Poisson manifold with the coordinates $X^I$ and a Poisson structure
$P^{IJ}(X)$, which satisfies the Jacobi identity
\begin{equation}
P^{IL}\partial_L P^{JK}+P^{KL}\partial_L P^{IJ}+P^{JL}\partial_L P^{KI}=0\,.\label{Jac}
\end{equation}
$A_{\mu I}$ is a gauge field. $\epsilon^{\mu\nu}$ is the antisymmetric Levi-Civita symbol.

All two-dimensional dilaton gravities \cite{Grumiller:2002nm} are particular cases of PSMs
corresponding to a three dimensional target space with $\{ A_{\mu I}\}=\{ \omega_\mu,e_{\mu a}\}$,
where $\omega_\mu$ is a spin-connection and $e_{\mu a}$ is a zweibein on the 2D space-time.
The target space coordinates $X^I$ are the dilaton field $X$ and two auxiliary fields $X^a$
that generate the torsion constraints.  It was demonstrated \cite{Kummer:1996hy}, that 2D
dilaton gravities are locally quantum trivial, which suggests that all dynamics in that
models should reside on the boundary.

An important result was obtained by Cattaneo and Felder \cite{Cattaneo:1999fm}, who demonstrated that
in a PSM on the disc the correlation functions of boundary values $X_{\rm b}^I$ of $X^I$ can be expressed through
the Kontsevich star product \cite{Kontsevich:1997vb}. In other words, the dynamics of $X_{\rm b}^I$ is a 
noncommutative quantum mechanics obtained by a quantization of the Poisson bracket
\begin{equation}
\{ X_{\rm b}^I,X_{\rm b}^I\} =P^{IJ}(X_{\rm b})\,. \label{CF}
\end{equation}

Here, we take the world-sheet to be a finite cylinder.
To analyze the model (\ref{PSM}) we use the Hamiltonian reduction method \cite{reduced}, i.e. before quantization
we construct a reduced phase space by solving the constraints and fixing the gauge freedom, see 
Sec.\ \ref{sec-Ham}. This reduced phase 
space appears to consist of the boundary values of $X^I$ and of the components $A_{r I}$ along the axis of
the cylinder. The reduced action contains a vanishing Hamiltonian, but has a non-trivial symplectic structure.
Quantization of such an action gives a noncommutative quantum mechanics, see Ref.\cite{NCQM} for some early
works. We apply the Fock space quantization scheme and use the Wick star product 
\cite{Waldmann} to calculate correlation
functions of the boundary fields (Sec.\ \ref{sec-quant}). Then, we proceed with the examples. The simplest
case of a constant Poisson tensor is briefly described in Sec.\ \ref{sec-const}. Linear Poisson structures
are considered in Sec.\ \ref{sec-lin}. For these latter models the reduced $0+1$ dimensional theories 
have (at least classically) a global symmetry group. In the case of the Jackiw-Bunster \cite{JT} model, this is the
conformal group $SO(2,1)$. For a generic Poisson structure, the correlation functions may be calculated as
perturbation series in the Poisson tensor $P^{IJ}$. These series are constructed in Sec\ \ref{sec-pert}. 

\section{Hamiltonian reduction}\label{sec-Ham}
Let us take the world sheet being a finite cylinder, $\mathcal{M}=S^1\times [0,l]$. Let $t$ denote a coordinate
on $S^1$ and $r\in [0,l]$. One has to impose some boundary conditions at $r=0,l$ that will ensure the absence
of boundary terms in the Euler-Lagrange variation and in the gauge transformation of the action (\ref{PSM}).
Such conditions are not unique. One possible choice is described below.
 
The gauge transformations
\begin{eqnarray}
&&\delta_\lambda X^I=P^{IJ}\lambda_J\,,\nonumber\\
&&\delta_\lambda A_{\mu I}=-\partial_\mu\lambda_I -\frac{\partial P^{JK}}{\partial X^I} \lambda_K A_{\mu J}
\label{PSMgauge}
\end{eqnarray}
leave the action (\ref{PSM}) invariant up to a total derivative,
\begin{equation}
\delta_\lambda S=\int_{\mathcal{M}} d^2\sigma \partial_\mu \left[ \epsilon^{\mu\nu} A_{\nu I}\lambda_J
\left( P^{IJ} -X^K \frac{\partial P^{IJ}}{\partial X^K} \right) \right]\,.\label{totder}
\end{equation}
This boundary term vanishes if we impose the boundary condition
\begin{equation}
A_{t I}|_{\partial\mathcal{M}}=0\,.\label{Abc}
\end{equation}
With this boundary condition, the Euler-Lagrange variations of (\ref{PSM}) do not produce
any boundary terms. The conditions (\ref{Abc}) are themselves gauge invariant if
\begin{equation}
\lambda|_{\partial\mathcal{M}}=0\,.\label{bclambda}
\end{equation} 

Our approach is pertubative, but not restricted to a finite order of the perturbation theory.
We assume, that the fields are "not too far" from the trivial background $A_{\mu I}=0=X^I$
(exactly as in \cite{Cattaneo:1999fm}). This will allow us to avoid the difficulties with 
non-existence of global gauge fixing conditions (Gribov ambiguities). We use the Hamiltonian
reduction method \cite{reduced}, that is especially well-suited for first-order theories.
Before quantizing, one has to fix the gauge freedom and solve the constraints. The latter
are generated by the fields $A_{t I}$ which play the role of Lagrange multiplies. The
constraints are
\begin{equation}
 \partial_r X^I+P^{IJ}A_{r J}=0 \label{cons}
\end{equation}

On the trivial background $A_{\mu I}=0=X^I$, the gauge transformations read: 
$\delta_\lambda A_{r I}=-\partial_r \lambda_I$. Taking into account the boundary condition
(\ref{bclambda}), one can easily see that the gauge freedom is fixed completely by 
\begin{equation}
A_{r I}(r,t)=a_I(t) \label{gfix}
\end{equation}
with some arbitrary functions $a_I(t)$. We shall assume that the condition (\ref{gfix}) still
selects a representative for each gauge orbit even in a vicinity of the trivial background.
Then, if the initial condition $x^I(t)=X^I(t,0)$ is in a sufficiently small region, the constraint
equation (\ref{cons}) has a unique solution $X(r;a_I(t),x^I(t))$. Therefore, the reduced phase
space variables are the boundary data $(a_I(t),x^I(t))$, and the action becomes ($\epsilon^{tr}=-1$)
\begin{equation}
 S_{\rm red}=-\int dt Y^I(a(t),x(t))\partial_t a_I(t)\,,\label{Sred}
\end{equation}
where
\begin{equation}
 Y^I(a,x):=\int_0^l dr\, X^I(r;a,x)\,.\label{YI}
\end{equation}
The action (\ref{Sred}) is $0+1$ dimensional, it gives a quantum mechanics upon quantization.
In terms of the $(Y^I,a_I)$ variables this model is trivial, but in terms of the interesting
variables $(x^I,a_I)$ (that are boundary values of the original fields $X^I$ and $A_{r I}$),
the action (\ref{Sred}) has a non-trivial symplectic structure, though the Hamiltonian vanishes.
The deformation quantization \cite{Waldmann} is probably the most appropriate method to quantize
such systems. The resulting quantum theory is a noncommutative quantum mechanics with a
position-dependent noncommutativity, see \cite{Gomes:2009tk} for examples of such theories.
The constraint equation (\ref{cons}) is non-linear. However, it can be solved perturbatively,
see Sec.\ \ref{sec-pert},
which is enough for our purposes.

To derive this result, we have assumed that there is a perhaps small but finite neighborhood
of the trivial background such that (\ref{gfix}) is an admissible gauge fixing and the constraints
have a unique solution for any initial value $x(t)$ from this neighborhood. Then our result
is valid to any order of the perturbative expansion around the trivial background (but is not
valid non-perturbatively beyond the neighborhood). Not all physically interesting PSMs
satisfy this assumption. In some cases, the Poisson tensor may be rather singular, see
Ref. \cite{Grumiller:2002nm}. However, for many reasonable PSMs this procedure does really
work, as we shall show below at several examples.

From a somewhat different perspective, the Hamiltonian analysis of PSMs was considered by Strobl
\cite{Strobl:1994yk}.

\section{Quantization}\label{sec-quant}
Let us consider the Fock space quantization of our system. This seems to be the most natural
choice of a quantization scheme, but not always the best one, as we shall briefly discuss at
the en of Sec.\ \ref{sec-lin}.
The canonical variables for the action (\ref{Sred}) with a generic Poisson structure are
\begin{equation}
q^I=Y^I,\qquad p_I=a_I\,.\label{qp}
\end{equation}
One can also introduce complex variables
\begin{equation}
z_I=\frac 1{\sqrt{2}} (q^I+ip_I)\,,\qquad \bar z_I=\frac 1{\sqrt{2}} (q^I-ip_I)
\label{zzbar}
\end{equation}
that upon quantization become creation and annihilation operators ${\bb}_I$ and ${\bb}_I^\dag$ with the
commutation relation
\begin{equation}
[\bb_I,\bb_J^\dag]=\delta_{IJ}\,.\label{combb}
\end{equation}
Note, that since the Hamiltonian is zero, the transition to Heisenberg time-dependent operators is trivial.

Let us stress again, that in terms of the variables (\ref{qp}) our quantum mechanical system for \emph{any}
Poisson structure is the most trivial one. However, these are not the variables one is interested in.
To analyze the holographic correspondence, one needs the correlation functions of the boundary values
$x^I,\, a_I$ of the original fields $X^I,\, A_I$. For a generic Poisson structure calculation of these
correlation functions is not trivial at all. Consider a function $f$ of $x$ and $a$. We associate to this function
a normal-ordered operator valued function $:{\bf f}:$ through the equation
\begin{eqnarray}
&&:{\bf f}:=\sum_{k,l}\frac 1{k!l!} \frac{\partial^{|k|+|l|}}{\partial_{z_1}^{k_1}\dots \partial_{z_n}^{k_n}
\partial_{\bar z_1}^{l_1}\dots \partial_{\bar z_n}^{l_n}} \, f\vert_{z=\bar z=0}\nonumber\\
&&\qquad\qquad\times \bb_1^{\dag l_1}\dots \bb_n^{\dag l_n}\bb_1^{k_1}\dots \bb_n^{k_n} \,.\label{norm}
\end{eqnarray}
$k$ and $l$ are multi-indices, $|k|=k_1+\dots +k_n$, $k!=k_1! k_2! \dots k_n!$.
The Fock vacuum is characterized by $\bb |0\rangle =0$. Therefore,
\begin{equation}
\langle :{\bf f}: \rangle \equiv  \langle 0| \, :{\bf f}:\, |0\rangle = f(z=\bar z=0) \,.\label{vacef}
\end{equation}
To calculate the vacuum expectation value of a product of several operator-valued functions, one has to
re-arrange this product in a normal-ordered form,
\begin{equation*}
:{\bf f}_1:\, :{\bf f}_2:\dots:{\bf f}_m:\, =\, :{\bf f}:\,.
\end{equation*}
The function $f$ on the right hand side of this equation is given by the Wick star product 
of the functions on the left hand side (see \cite{Waldmann}),
\begin{equation}
f=f_1\star f_2 \star \dots \star f_m \,,\label{fff}
\end{equation}
that reads
\begin{eqnarray}
&&f\star g = f \exp(\overleftarrow{\partial}_{z_K}\overrightarrow{\partial}_{\bar z_K})\, g\label{star}\\
&&=f\cdot g +\partial_{z_K}f \cdot \partial_{\bar z_K}g
+\frac 12 \partial_{z_K}\partial_{z_L}f \cdot \partial_{\bar z_K}\partial_{\bar z_L}g+\dots
\nonumber
\end{eqnarray}

The problem of calculation of correlation functions of the boundary data $x^I$, $a_I$ has been reduced to
calculations of the derivatives with respect to $z$ and $\bar z$. To this end, we have to invert the formulas
(\ref{cons}) and (\ref{YI}) to express $x^I$ though the canonical variables $p$ and $q$.

\section{Constant Poisson structure}\label{sec-const}
As a warm-up, let us take
\begin{equation}
P^{IJ}=\frac 2l w^{IJ}\,,\label{constP}
\end{equation}
where $w^{IJ}$ is a constant antisymmetric matrix. Then, the gauge transformation (\ref{PSMgauge})
of $A_{\mu I}$ is just the gradient transformation, so that the condition (\ref{gfix}) is an allowed
gauge fixing on the whole phase space. The constraints (\ref{cons}) have a unique solution for
arbitrary initial data
\begin{equation}
X^I(r;a,x)=x^I(t)-\frac{2r}l w^{IJ}a_J(t) \label{wsol}
\end{equation}
and
\begin{eqnarray}
&&Y^I(a,x)=l(x^I-w^{IJ}a_J)\,,\label{wY}\\
&&S_{\rm red}=-l\int dt (x^I-w^{IJ}a_J)\,\partial_t a_I \,.\label{wSred}
\end{eqnarray}
The transformation of the canonical coordinates $x^I\to Y^I/l=x^I-w^{IJ}a_J$ is nothing else than the
famous Bopp shift, that relates noncommutative theories with a constant noncommutativity to commutative
ones.

The boundary fields depend linearly on the canonical variables,
\begin{equation}
a_I=p_I,\qquad x^I=\frac {q^I}l + w^{IJ}p_J\,,\label{constinv}
\end{equation}
which facilitates calculations of the star product (\ref{star}) and of the correlation functions.

\section{Linear Poisson structure}\label{sec-lin}
Let us take a linear Poisson structure,
\begin{equation}
P^{IJ}(X)=C^{IJ}_K X^K\,.\label{linP}
\end{equation}
Due to the Jacobi identity on $P^{IJ}$, the constants $C^{IJ}_K$ have to be structure
constants of a Lie algebra of some Lie group $\mathcal{G}$. The gauge transformations 
(\ref{PSMgauge}) become just the
usual Yang-Mills type transformations, though the gauge group need not be compact in
our case. If the gauge group is $\mathcal{G}=SO(2,1)$ the corresponding PSM is nothing else than the
Jackiw-Bunster gravity \cite{JT}.

The dimension of the space of gauge orbits is locally constant. Therefore,
to prove that the gauge condition (\ref{gfix}) is admissible, it is enough to check
that different functions $a^I$ belong to different gauge orbits. We check an infinitesimal
version of this statement. Suppose that a constant $a^I$ and a constant $a^I + \delta a^I$
are related through a gauge transformation (we omit the $t$-dependence that is not essential
here), i.e,
\begin{equation}
\partial_r \lambda =-\widehat{Ca}\lambda -\delta a \,,\label{linl}
\end{equation}
where we suppressed the indices and introduced a matrix $(\widehat{Ca})^K_I\equiv C^{JK}_Ia_J$.
One can easily find the function $\lambda(r)$ that satisfies (\ref{linl}) and the Dirichlet
boundary condition at $r=0$. It reads
\begin{equation}
\lambda (r)=\exp(-r\, \widehat{Ca}) \int_0^r d\rho \, \exp(\rho\, \widehat{Ca})\delta a\,.
\end{equation}
Next, we have to  check whether there is a choice of $\delta a$ such that the other boundary
condition, $\lambda (l)=0$, is also satisfied. By a (possibly complex) change of the basis
one can bring $\widehat{Ca}$ to the canonical Jordan form. Clearly, different Jordan blocks
may be considered separately. Let us take one of these blocks,
\begin{equation}
(\widehat{Ca})_1=
\left( \begin{array}{ccccc} z_0 & z_1 & 0 & \cdots & 0 \\
                           0 & z_0 & z_1 & \cdots & 0 \\
                           \vdots & \ddots & \ddots & \ddots & \vdots \\
                           0 & \cdots & 0 & z_0 & z_1 \\
                           0 & \cdots & 0 & 0 & z_0 \end{array} \right)\label{block}
\end{equation}
with some numbers $z_0$ and $z_1$ depending linearly on $a_I$. An easy computation shows
\begin{equation}
\det \int_0^l d\rho \, \exp(\rho\, (\widehat{Ca})_1)=\left[ \frac{ e^{z_0l} -1}{z_0} \right]^n \,,
\end{equation}
where $n$ is the dimension of the block.
This determinant can be zero only if $z_0$ is pure imaginary, i.e. if the gauge group has a
compact subgroup. Even in this case, for sufficiently small $z_0$ (equivalently, for sufficiently
small $a^I$) the determinant above is non-zero. Therefore, there is a neighborhood of the trivial
vacuum, such that (\ref{linl}) has no solutions, and the gauge condition (\ref{gfix}) is admissible.

In the index-free notations the constraint equation (\ref{cons}) reads $\partial_r X -(\widehat{Ca})^TX=0$,
where the transposition means that the index of $X$ is now contracted with the lower index of $\widehat{Ca}$
(instead of the upper index in (\ref{linl})). This equation has a unique solution for any initial data,
\begin{equation}
X(r;a,x)=\exp \big( r(\widehat{Ca})^T\big) x \,,\label{linXx}
\end{equation}
so that the reduced action becomes
\begin{equation}
S_{\rm red}=-\int dt \left[ \int_0^l dr \exp \big( r(\widehat{Ca})^T\big) x \right]^I \partial_t a_I\,.
\label{linS}
\end{equation}
This result admits a geometric interpretation. For example, the solution (\ref{linXx}) is a parallel
transport of the initial value $x^I$ by a one-parameter group generated by $(\widehat{Ca})^T$.

For a constant parameter, $\lambda_I(r,t)=\varepsilon_I$, the transformations (\ref{PSMgauge}) read
\begin{equation}
\delta_\varepsilon x=-(\widehat{C\varepsilon})^Tx,\qquad \delta_\varepsilon a=\widehat{C\varepsilon}\, a
\,.\label{globxa}
\end{equation}
Under these transformations,
\begin{equation}
\delta_\varepsilon \widehat{Ca}=[\widehat{C\varepsilon},\widehat{Ca}],\qquad
\delta_\varepsilon Y=-(\widehat{C\varepsilon})^T Y\,,\label{globCaY}
\end{equation}
so that the action (\ref{linS}) remains invariant. Therefore, holographic duals to Poisson
sigma models with a linear Poisson structure have a global symmetry group $\mathcal{G}$. In the case of the
Jackiw-Bunster gravity \cite{JT} this group is $SO(2,1)$, which is is the conformal group in one
dimension. Thus, the holographic dual to the Jackiw-Bunster gravity is a conformal noncommutative
quantum mechanics. 

In general, the relation (\ref{zzbar}) breaks the $\mathcal{G}$-invariance. One should rather use the
expressions like $ip_I+g_{IJ}q^I$, where $g_{IJ}$ is the matrix of a $\mathcal{G}$ invariant scalar product
on the Lie algebra. If $\mathcal{G}$ is compact, there is a positive-definite invariant scalar product,
so that after a suitable re-scaling one can take $g_{IJ}=\delta_{IJ}$. The structure constants $C^{IJ}_K$
become totally antisymmetric. One can use the quantization scheme of Sec.\ \ref{sec-quant}. The Fock vacuum
is $\mathcal{G}$-invariant. Although (\ref{linXx}) is highly non-linear, one can express the boundary data
through the canonical variables in a closed form: $a=p$, $x=G^{-1}(p)q$, where
\begin{equation}
G(p)=\int_0^l dr \exp(r (\widehat{Cp})^T)\,.\label{Gp}
\end{equation}
This is enough to calculate the star products. The calculations of correlators are simplified by the fact
that one has to put all canonical variables equal to zero after computing the products.

Unfortunately, this simple quantization scheme does not work for non-compact groups, since they do not admit a
positive definite invariant bilinear form. The Fock vacuum fails to be $\mathcal{G}$-invariant. As was
demonstrated in \cite{AdS2QM} at the example of de~Alfaro-Fubini-Furlan \cite{de Alfaro:1976je} 
quantum mechanics, one has to use a different quantization scheme. Correct conformal correlation
functions are recovered through a rather sophisticated procedure. Since the analysis of \cite{AdS2QM}
was based mostly on the $SO(2,1)$ group structure, one may probably apply it to the holographic dual
of Jackiw-Bunster gravity as well, which we are going to do in the future. 
\section{Perturbative expansions}\label{sec-pert}
For a generic Poisson structure, the best one can do is to solve the constraints and 
calculate the correlation function in the form of a perturbative expansion in $P^{IJ}$. Let us rescale
$P^{IJ}\to \theta P^{IJ}$. A formal parameter $\theta$ is introduced to count the orders of
perturbation theory only. One has to put $\theta=1$ at the end of the calculations. Let us
expand a solution of the constraint equation (\ref{cons}) (that now reads
$\partial_r X^I+\theta P^{IJ}(X)a_{J}=0$) as
\begin{equation}
X^I=X^I_0+\theta X^I_1+\theta^2 X^I_2 +\dots \label{XIpert}
\end{equation}
The Poisson tensor is also expanded,
\begin{eqnarray}
&&P^{IJ}(X)=P^{IJ}(X_0)+\theta \partial_K P^{IJ}(X_0)X_1^K + \label{Ppert}\\
&&+ \theta^2\bigl( \partial_K P^{IJ}(X_0) X_2^K +\tfrac 12 \partial_K \partial_L
P^{IJ}(X_0) X_1^KX_1^L\bigr) + \dots \nonumber
\end{eqnarray}
The initial conditions read
\begin{equation}
X^I_0(0)=x^I,\quad X^I_k(0)=0,\ k=1,2,\dots \label{ini}
\end{equation}
The constraint equation may be easily solved yielding
\begin{eqnarray}
&&X^I_0=x^I,\nonumber\\
&&X^I_1=-r P^{IJ}(x)a_J\,,\nonumber\\
&&X^I_2=\frac {r^2}2 \, \partial_KP^{IJ}(x)\cdot P^{KL}(x)a_Ja_L\,.
\label{XI012}
\end{eqnarray}
Next, we integrate this solution over $r$ (see (\ref{YI})) to obtain
\begin{eqnarray}
&&Y^I(a,x)=lx^I-\frac {l^2}2 \theta P^{IJ}(x)a_J\nonumber\\
&&\quad +\frac {l^3}6 \theta^2 \partial_KP^{IJ}(x)\cdot P^{KL}(x)a_Ja_L 
+O(\theta^3)\label{YIp}
\end{eqnarray}
One has to express $x^I$ through the canonical variables $q^I=Y^I$ and $p_I=a_I$.
\begin{eqnarray}
&&x^I=\frac {Y^I}l+\theta \frac l2 P^{IJ}(Y/l) \nonumber\\
&&+\frac {l^2}3 \theta^2 \partial_KP^{IJ}(Y/l)\cdot P^{KL}(Y/l)a_Ja_L 
+O(\theta^3)\label{xIp}
\end{eqnarray}

Now, we are ready to calculate the correlation functions. For example,
\begin{eqnarray}
&&\langle :x^I:\, :x^K: - :x^K:\, :x^I:\rangle \nonumber\\
&&\quad = (x^I\star x^K-x^K\star x^I)\vert_{q=p=0}= -iP^{IK}(0) \label{xIxK}
\end{eqnarray}
where we restricted ourselves to the order $\theta^1$ and put $\theta=1$ after the
calculation. The relation (\ref{xIxK}) reminds us of Eq.\ (\ref{CF}).

\section{Conclusions}
We have demonstrated that if the Poisson structure behaves sufficiently well, i.e. if
there is a perhaps small but finite vicinity of the trivial vacuum such that 
the gauge condition (\ref{gfix}) is admissible and the constraint (\ref{cons}) has a
unique solution, the corresponding Poisson sigma model is (perturbatively) equivalent 
to a noncommutative quantum mechanics for the boundary data. The example of a linear
$P^{IJ}$ shows that regularity of the Poisson structure is not needed.

The present work can be viewed as a extension of the results of \cite{Cattaneo:1999fm} from a disc to
a finite cylinder.
The topology of a cylinder is a natural arena for the Hamiltonian reduction method.
Therefore, we were able to extend the noncommutative description to all fields of the model.
It would be interesting and important to check our results with other quantization
methods, as the ones used in \cite{Cattaneo:1999fm}. 

Fock space quantization and the Wick star product provide us with an effective way to calculate
the correlation functions for boundary even though for a generic Poisson structure one has
to use a perturbative expansion in $P^{IJ}$. 

The most explicit but yet nontrivial results are obtained for linear Poisson structures. In this
models, the Hamiltonian reduction yields a classical $0+1$ dimensional action which is 
invariant under a global symmetry group. For the Jackiw-Bunster model \cite{JT}, this is the conformal
group $SO(2,1)$. Although our quantization methods are, strictly speaking, not applicable to
the cases with non-compact symmetry groups, our results support the point of view that a
conformal quantum mechanics may be realized as a holographic dual to a PSM.

Our results are not immediately applicable  to two-dimensional gravities since the trivial
vacuum corresponds to a degenerate metric and is not a natural expansion point. Besides,
in gravity models one is usually interested in asymptotic conditions at the conformal boundary
of AdS$_2$ rather than in the conditions at a finite boundary. Although we got a quantum mechanics
as a holographic dual, we cannot exclude that for the AdS gravity case the dual theory will be
a chiral half of a CFT. The methods
suggested here can definitely be extended to AdS gravity models as well. 

\begin{acknowledgments}
I am grateful to Daniel Grumiller for numerous discussions on low-dimensional holography and
to Vlad Kupriyanov for useful comments.
This work was supported in part by FAPESP and CNPq.
\end{acknowledgments}

\end{document}